\documentclass[twocolumn,showpacs,aps,prl,floatfix,superscriptaddress,letterpaper]{revtex4}

\usepackage{graphicx}
\usepackage{dcolumn}
\usepackage{amsmath}
\usepackage{epsfig}

\input babarsym

\newcommand{\BABARPubYear}    {07}
\newcommand{\BABARPubNumber}  {051}

\newcommand{\SLACPubNumber} {12713}

\newcommand {\Bsg}    {\ensuremath{\B \rightarrow X_s \gamma}\xspace}
\newcommand {\Bxlnu}  {\ensuremath{\Bbar \rightarrow X \ell \bar{\nu}}\xspace}
\newcommand {\Bxclnu} {\ensuremath{\Bbar \rightarrow X_c \ell \bar{\nu}}\xspace}
\newcommand {\Bxulnu} {\ensuremath{\Bbar \rightarrow X_u \ell \bar{\nu}}\xspace}

\newcommand {\mx}     {\ensuremath{M_{X}}\xspace}
\newcommand {\Pplus}  {\ensuremath{P_{+}}\xspace}
\newcommand {\Q}      {\ensuremath{q^{2}}\xspace}

\newcommand {\mX}     {\ensuremath{M_{X}}\xspace}
\newcommand {\mmiss}  {\ensuremath{m_{miss}^2}\xspace}

\newcommand {\breco}  {\ensuremath{B_\mathrm{reco}}\xspace}
\newcommand {\brecoil}{\ensuremath{B_\mathrm{recoil}}\xspace}

\newcommand{\btoxlnu}{\ensuremath{\Bbar\to X\ell\overline{\nu}_{\ell}}}
\newcommand {\rusl}   {\ensuremath{R_\mathrm{u/sl}}}

\newcommand{\GEVCC}{\ensuremath{{\mathrm{\,Ge\kern -0.1em V^2\!/}c^4}}\xspace}
\newcommand{\GEV}{\ensuremath{{\mathrm{\,Ge\kern -0.1em V^2}}}\xspace}
\newcommand{\GEVC}{\ensuremath{{\mathrm{\,Ge\kern -0.1em V/}c}}\xspace}
\newcommand{\gevccsq}{\ensuremath{{\mathrm{\,Ge\kern -0.1em V^2\!/}c^4}}\xspace}

\newcommand{\rb}[1]{\raisebox{1.5ex}[0pt]{#1}}
\newcommand{\rbd}[1]{\raisebox{-0.4ex}[0pt]{#1}}
\newcommand{\rbu}[1]{\raisebox{0.4ex}[0pt]{#1}}

\begin{document}

\preprint{\babar-PUB-\BABARPubYear/\BABARPubNumber} 
\preprint{SLAC-PUB-\SLACPubNumber} 

\begin{flushleft}
\babar-PUB-\BABARPubYear/\BABARPubNumber\\
SLAC-PUB-\SLACPubNumber\\
\end{flushleft}

\title {\large\bf Measurements of Partial Branching Fractions for
{\boldmath $\Bxulnu$} and Determination of {\boldmath $\Vub$}}

%
\author{B.~Aubert}
\author{M.~Bona}
\author{D.~Boutigny}
\author{Y.~Karyotakis}
\author{J.~P.~Lees}
\author{V.~Poireau}
\author{X.~Prudent}
\author{V.~Tisserand}
\author{A.~Zghiche}
\affiliation{Laboratoire de Physique des Particules, IN2P3/CNRS et Universit\'e de Savoie, F-74941 Annecy-Le-Vieux, France }
\author{J.~Garra~Tico}
\author{E.~Grauges}
\affiliation{Universitat de Barcelona, Facultat de Fisica, Departament ECM, E-08028 Barcelona, Spain }
% ...
\collaboration{The \babar\ Collaboration}
\noaffiliation

\date{\today}

\begin{abstract}
We present partial branching fractions for inclusive charmless semileptonic $B$
decays $\kern 0.18em\overline{\kern -0.18em B}{} \rightarrow X_u \ell \bar{\nu}$, and the determination of the CKM matrix element $|V_{ub}|$.
The analysis is based on a sample of 383 million $\Upsilon{(4S)}$ decays into 
$\ensuremath{B\kern 0.18em\overline{\kern -0.18em B}{}\xspace}$
pairs
collected with the $\mbox{\slshape B\kern-0.1em{\smaller A}\kern-0.1em B\kern-0.1em{\smaller A\kern-0.2em R}}$ detector at the 
PEP-II $e^{+} e^{-}$ storage rings. 
We select events using either the invariant mass $M_{X}$ of the hadronic system, 
the invariant mass squared, $q^{2}$, of the lepton and neutrino pair, the kinematic variable
$P_{+}$ or one of their combinations. We then determine partial branching fractions in limited regions of phase space:
$\Delta {\cal{B}} =(1.18 \pm 0.09_{\rm stat.} \pm 0.07_{\rm syst.} \pm 0.01_{\rm theo.}) \times 10^{-3}$ 
($M_{X}<1.55~{\rm GeV}/c^{2}$),
$\Delta {\cal{B}} =(0.95 \pm 0.10_{\rm stat.} \pm 0.08_{\rm syst.} \pm 0.01_{\rm theo.}) \times 10^{-3}$ 
($P_{+} < 0.66~{\rm GeV}/c$), and
$\Delta {\cal{B}} =(0.81 \pm 0.08_{\rm stat.} \pm 0.07_{\rm syst.} \pm 0.02_{\rm theo.}) \times 10^{-3}$ 
($M_{X}<1.7~{\rm GeV}/c^{2}$, $q^{2}>8~{\rm GeV}^{2}/c^{4}$).
Corresponding values of $|V_{ub}|$ are extracted using several theoretical calculations.
\end{abstract}

\maketitle


In the Standard Model the element $V_{ub}$ of the Cabibbo-Kobayashi-Maskawa (CKM) quark-mixing
matrix~\cite{CKM} plays a critical role in tests of the prediction of 
$\CP$ violation. 
Since the rate for charmless semileptonic decays, $\Bxulnu$~\cite{hadronicsystem}, 
is proportional to $\Vub^2$, and the hadronic and leptonic currents are factorizable, the best 
method to extract this quantity is to measure branching fractions for such decays~\cite{vub2004}.  
Experimentally, the principal challenge is to separate the rare $\Bxulnu$ decays from the approximately 
$50$ times  larger $\Bxclnu$ background.  
Given that the $u$ quark is much lighter than the $c$ quark, regions of phase space 
can be defined where the background is suppressed.
To relate the decay rate of the \B\ meson to $\Vub$, parton level calculations have to be 
corrected for perturbative and non-perturbative QCD effects.  
A variety of QCD calculations are available to determine these corrections~\cite{blnp,dge,bll}. 

In this letter, we present a measurement of partial branching
fractions for inclusive charmless semileptonic decays, 
$\Bxulnu$~\cite{chconj}.
$\FourS \to \BB$ events are tagged by the full reconstruction of a hadronic decay 
of one of the \B\ mesons (\breco). The semileptonic decay of the second \B\ meson 
(\brecoil) is identified by the presence of an electron or a muon.
This technique results 
in a low event selection efficiency but allows the determination of the 
momentum, charge, and flavor of the \B\ mesons. 

We use three kinematic variables to separate \Bxulnu\ decays from the 
dominant \Bxclnu\ background: \mx, the invariant mass of the hadronic system
$X_{u,c}$; \Q, the invariant mass squared of the lepton-neutrino system; 
and 
$\Pplus \equiv E_X-|\vec{P}_X|$~\cite{blnp,dge}, 
where $E_X$ and $\vec{P}_X$ are the energy and momentum 
of the hadronic system $X_{u,c}$ calculated in the \B\ rest frame.
We measure the fraction of partial rates of 
charmless semileptonic 
decays $\Delta \rusl={\Delta\BR(\Bxulnu)/\BR(\Bxlnu)}$ in restricted 
phase space regions, corrected for resolution effects.
The resulting partial branching fractions are used to calculate 
$\Vub$ following theoretical prescriptions.

The analysis uses a sample of 383 million \FourS\ decays into \BB\
pairs, corresponding to an integrated luminosity of $347.4 ~\rm fb^{-1}$, collected with the \babar\ detector~\cite{detector}. 
Charmless semileptonic \Bxulnu\ decays are simulated as a combination 
of three-body decays ($X_u = \pi, \eta, \etapr, 
\rho, \omega, \ldots$)~\cite{ref:isgwtwo} and decays to non-resonant hadronic
final states $X_u$~\cite{ref:fazioneubert}. The motion of the $b$ quark inside 
the \B\ meson is modeled with the shape function parametrization given 
in Ref.~\cite{ref:fazioneubert}.
The simulation of the \Bxclnu\ background uses an HQET parametrization of 
form factors for $\Btodstlnu$~\cite{ref:caprini,ref:dstarlnuffbabar}, and models for 
$\Bb\to D \pi \ell\nub, D^* \pi \ell\nub$~\cite{ref:goityroberts}, and for 
$\Bb\to D \ell\nub,D^{**}\ell\nub$~\cite{ref:isgwtwo}.
The simulation of the hadronization is 
performed by \jetset74 ~\cite{ref:jetset}.
We use GEANT4~\cite{geant} to simulate the detector response.

To reconstruct a large sample of hadronically decaying \B\ mesons, 
$\breco \rightarrow  \Db^{(*)} Y^{\pm}$ are selected. 
Here, the system $Y^{\pm}$ consists of hadrons with a total charge of 
$\pm 1$, composed of $n_1\pi^{\pm}\, n_2K^{\pm}\, n_3\KS\,  n_4\piz$, 
where $n_1 + n_2 \leq 5$,  $n_3  \leq  2$,  and  $n_4  \leq  2$.   
The kinematic consistency of $\breco$ candidates 
is checked with two variables, 
$\mes = \sqrt{s/4 - \vec{p}^{\,2}_B}$ and 
$\Delta E = E_B - \sqrt{s}/2$. Here $\sqrt{s}$ is the total
energy in the \FourS\ center of mass frame, and $\vec{p}_\B$ and $E_\B$
denote the momentum and energy of the $\breco$ candidate in the same
frame. We require $\Delta E = 0$ within three standard deviations as 
measured for each decay mode. 
For each of the \breco\ decay modes, the purity $\cal{P}$ is estimated
using Monte Carlo (MC) simulation. $\cal{P}$ is defined as the ratio of signal over background events 
with $\mes \geq 5.27 \gevcc$. Only
modes for which $\cal{P}$ exceeds 20\%\ are used. On average, 
we reconstruct at least one \B\ candidate in 0.3\%  (0.5\%) of the \BzBzb\ 
(\BpBm) events. For events with more than one reconstructed \B\ 
decay, the decay mode with the highest purity is selected.

\begin{figure}[t]
    \begin{centering}
      \includegraphics[width=0.5\textwidth]{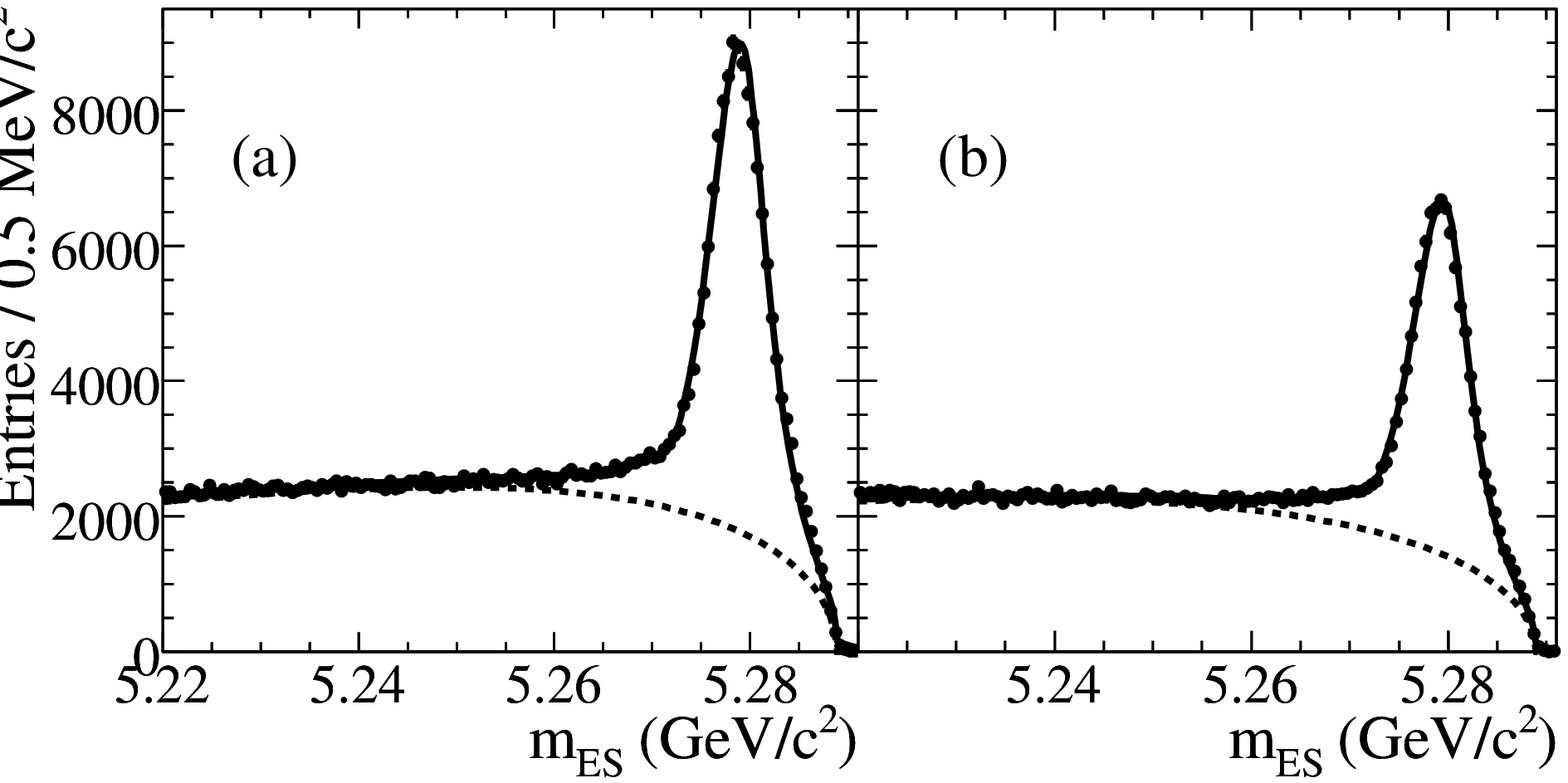}
      \caption{The \mes\ distribution for data (full circles) is shown together with the results of 
      the fit (solid line) for selected semileptonic decays from \BpBm\ events (a) and \BzBzb\ events (b). The dashed 
      line shows the contribution from combinatorial and continuum background.}
    \label{fig:msl}
   \end{centering}
   \end{figure}

We determine the number of \breco\ candidates from an unbinned maximum 
likelihood fit to the \mes\ distribution.
The data are fit to the sum of three contributions: signal $\breco$ decays, combinatorial 
background from \BB\ events, and continuum ($e^+e^- \to q\bar{q}$, $q = \u,\d,\s,\c$) events.
A Threshold function~\cite{Argus} is used to 
describe the combinatorial and continuum backgrounds.
To obtain a good description of the signal \mes\ distribution,
we adopt the modified Gaussian function used in Ref.~\cite{thorsten},
to account for energy losses of photons in the detector.
Fits to the \mes\ distribution are shown in Fig.~\ref{fig:msl}.
Semileptonic decays $\Bxlnu$ of the $B_{\rm recoil}$ candidate are identified by an electron or muon 
with momentum, $p^*_{\ell}$, defined in the \Bb rest frame, greater than 1~\gevc. For charged \breco\ candidates, 
we require the charge of the lepton to be consistent with a prompt 
semileptonic \Bb\ decay. For neutral \breco\ candidates, both charge-flavor
combinations are retained and the known average $\Bz$-$\Bzb$ mixing rate~\cite{PDG2006} 
is used to extract the prompt lepton yield.

The hadronic system $X$ in the decay \Bxlnu\ is reconstructed from charged
tracks and energy depositions in the calorimeter that are not associated 
with the \breco\ candidate or the identified lepton. 
We reconstruct \KS by performing a mass-constrained fit to $\pi^+ \pi^-$ pairs with
an invariant mass in the range 0.473-- 0.523~\gevcc.
The neutrino four-momentum $p_{\nu}$ is estimated from the missing 
momentum four-vector $p_{\mathrm{miss}} = p_{\Upsilon(4S)}-p_{\breco} -p_X-p_\ell$, 
where all momenta are measured in the laboratory frame and $p_{\Upsilon(4S)}$ 
refers to the \FourS\ meson. 

To select \Bxulnu\ candidates we require exactly one charged lepton with 
$p^*_{\ell} > 1 \gevc$, charge conservation ($Q_{X} +Q_\ell + Q_{\breco} = 0$), 
and a missing mass consistent with zero ($\mmiss < 0.5 \gev^2/c^4$).  
These criteria suppress the dominant \Bxclnu\ decays, many of which 
contain additional leptons or an undetected $\KL$ meson. We suppress
the $\Btodstlnu$ background by reconstructing
the low momentum $\pi^+$ from the $\Dstarp\to \Dz\pi^+$ decay. 
Since the momentum of the $\pi^+$ is almost collinear 
with the \Dstarp\ momentum $p_{\Dstarp}$, we can approximate 
the \Dstarp\ energy as $E_{\Dstarp}
\simeq m_{\Dstarp} \times E_{\pi} /145 \mevcc$. 
The neutrino mass $m_{\mathrm{veto}}^2 = (p_B - p_{\Dstarp} - p_{\ell})^2$ is peaked at zero 
for background events. The requirement $m_{\mathrm{veto}}^2 < - 3\gev^2/c^4$ reduces the 
$\Btodstlnu$ background by about 36\% while keeping more than 90\% of signal events.
We reject events with charged kaons or \KS\ in the  \brecoil  to reduce  
the background from \Bxclnu\ decays.    
   \begin{figure*}[htbp]
    \begin{centering}
      \includegraphics[width=0.25\textwidth,totalheight=7cm]{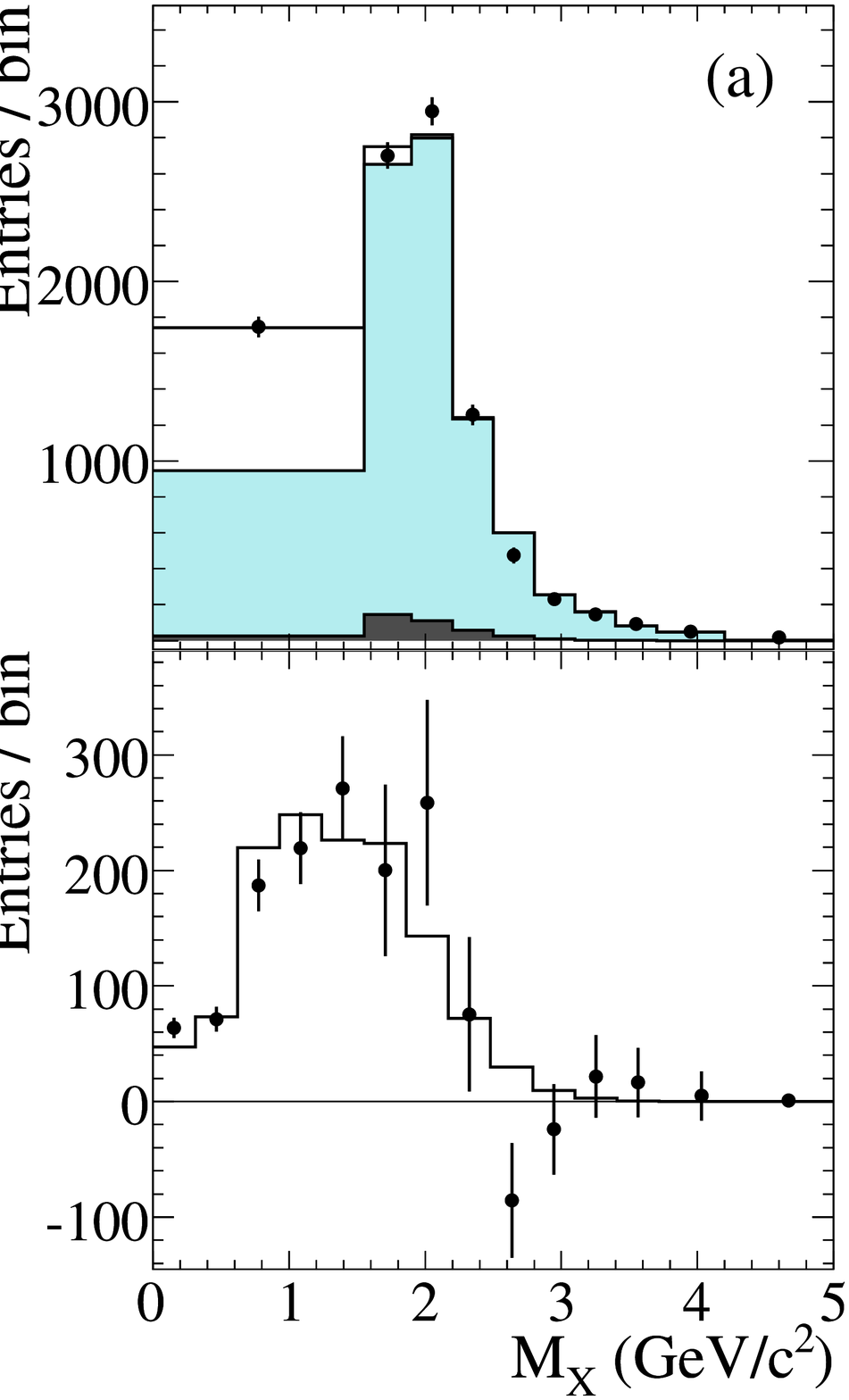}
      \includegraphics[width=0.25\textwidth,totalheight=7cm]{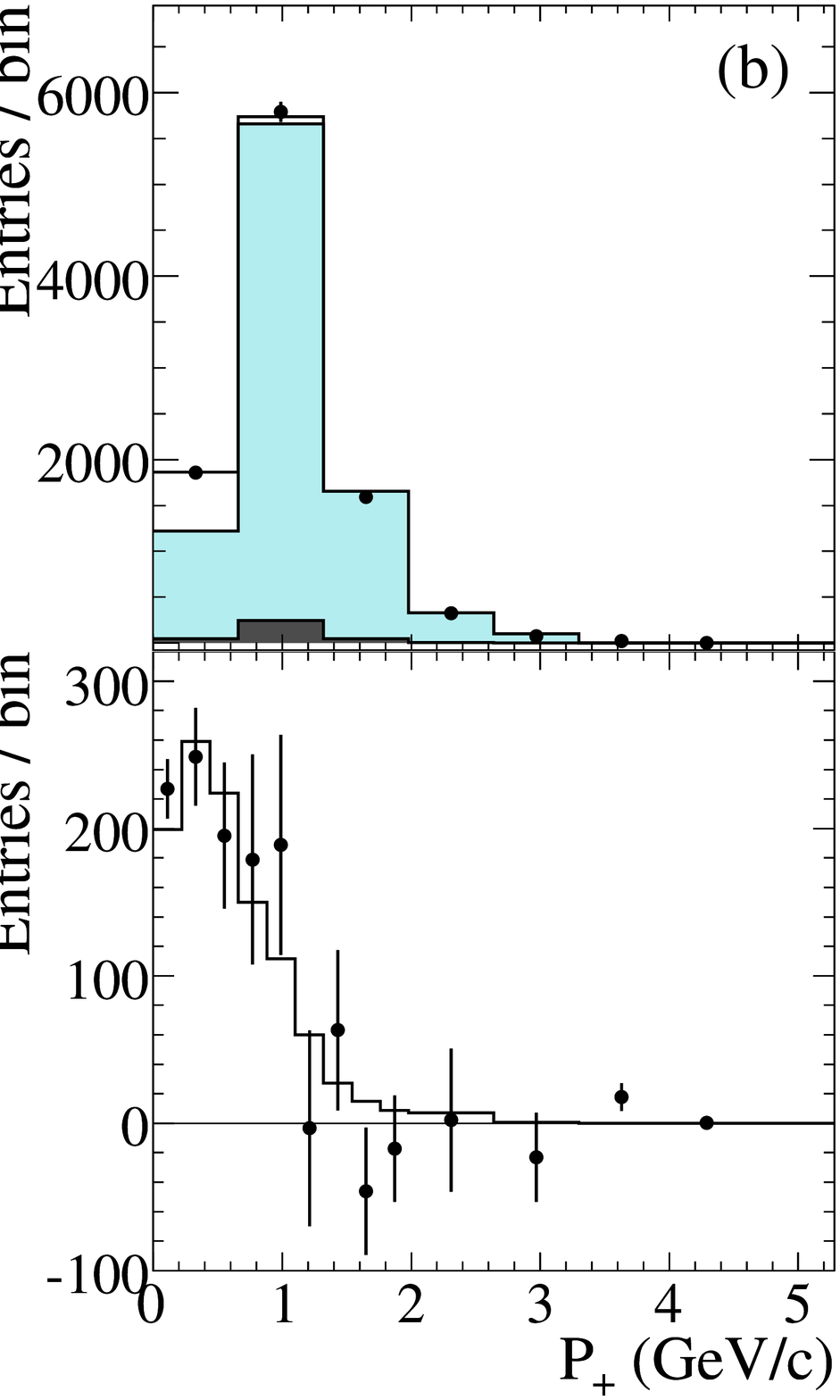}
      \includegraphics[width=0.25\textwidth,totalheight=7cm]{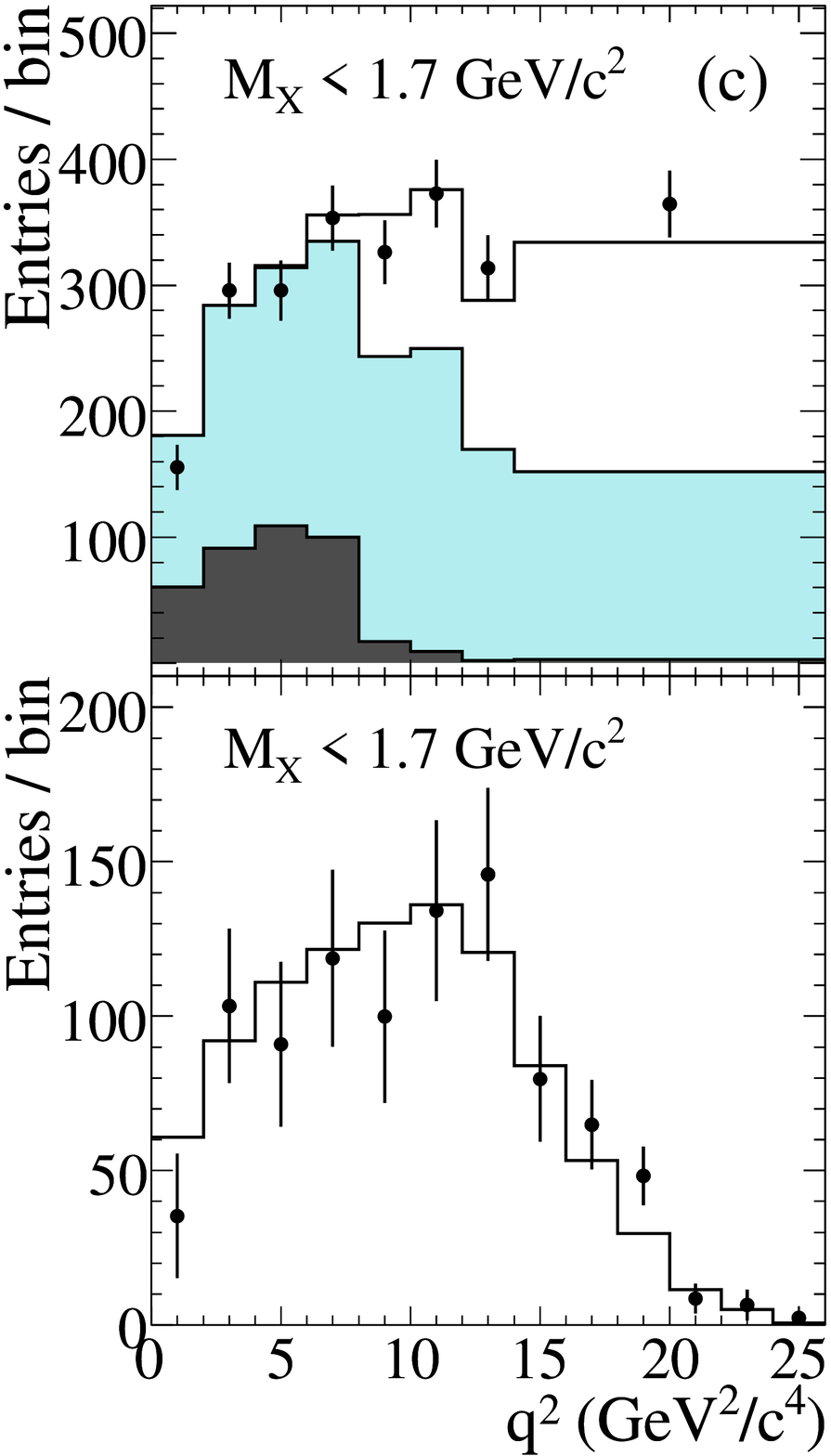}
      \caption{Upper row: measured \mx\ (a), \Pplus\ (b) and \Q\ with $\mx<1.7~\gevcc$ (c)
	spectra (data points). The result of the fit to the sum of three MC contributions is shown 
        in the histograms: 
	\Bxulnu\ decays generated inside (no shading) and outside (dark shading) the 
	selected kinematic region, and \Bxclnu\ and other background (light shading).  
        Lower row: corresponding spectra for \Bxulnu\ 
        after \Bxclnu\ and other background subtraction;
        they have been rebinned in order to show the shape of the kinematic variables.}
    \label{fig:mxplots}
   \end{centering}
   \end{figure*}

To extract the distribution in the variables \mx, \Pplus, and  the combination of \mx\ and \Q,
we perform fits to the \breco\ \mes\ distributions for subsamples of events in individual bins for each of the variables, 
and subsequently separate the signal from the combinatorial and continuum backgrounds for the three distributions. 
The resulting distributions are presented in Fig.~\ref{fig:mxplots}.
To reduce the systematic uncertainties in the derivation of the branching fractions we determine 
the ratios of the partial branching fractions to the total semileptonic branching fraction.
This is done for restricted regions of phase space, $\mx<1.55~\gevcc$, $\Pplus<0.66~\gevc$,
and ($\mx<1.7~\gevcc$, $\Q>8.0~\gevccsq$).
Specifically we define this ratio as
 \begin{equation}
 \frac{\Delta {\ensuremath{\cal B}\xspace}(X_u \ell \bar{\nu}_\ell)}{{\ensuremath{\cal B}\xspace}(X \ell \bar{\nu}_\ell)}=
 \frac{(N_u-N_u^{\rm{out}}-BG_u)/(\epsilon_\mathrm{sel}^u \epsilon_\mathrm{kin}^u)}
      {(N_\mathrm{sl}-BG_\mathrm{sl})}
 \times \frac{\epsilon_\ell^\mathrm{sl} \epsilon_t^\mathrm{sl} } {\epsilon_\ell^u \epsilon_t^u },
 \label{eq:ratioBR}
 \end{equation}
where $N_u$ refers to the number of observed events, $BG_u$ 
to the estimated number of background events, and $N_u^{\rm{out}}$ to the signal events that migrate
from outside the kinematic region into the signal region.
They are determined by a $\chi^{2}$ fit to the measured spectra with signal and background shapes
determined from MC simulation. 
$N_\mathrm{sl} = 181074 \pm 706$ and $BG_\mathrm{sl} = 12185 \pm 78$ are the number of semileptonic events,
extracted with a \mes\ fit, and the
corresponding background, determined from simulation.
The efficiency $\epsilon_\mathrm{sel}^u$ denotes the fraction of selected \breco-tagged signal events 
with a high-energy lepton.
The model-dependent efficiency $\epsilon_\mathrm{kin}^u$ accounts for the loss of selected events 
generated in the kinematic region that migrate outside this region.
The efficiency of the tag and lepton selection, $\epsilon_t$ and $\epsilon_\ell$, differ slightly for the
signal and the semileptonic samples, due to differences in the lepton momentum distribution 
and the multiplicity of the recoiling \B\ meson. 
To convert the ratio in Eq.~\ref{eq:ratioBR} to partial branching fractions, 
we use the total semileptonic branching fraction,
$\BR(\btoxlnu)= (10.75 \pm 0.15)\%$~\cite{PDG2006}.
The resulting partial branching fractions for the three selected kinematic regions, along with 
parameters in Eq.~\ref{eq:ratioBR},
are listed in Table~\ref{tab:inputdeltaB}. 
The statistical correlations between the \mx\ and (\mx,\Q), 
\Pplus\ analyses are 65\%, 67\%, 38\% respectively.

\begin{table*}[htbp]
\begin{center}
\caption{Summary of the fitted number of events and efficiencies, $\Delta \BR(\Bxulnu)$, and extracted $\Vub$
for the three kinematic cuts. The first uncertainty is statistical, the second systematic.
For $\Delta \BR$, the third uncertainty is due to the theoretical knowledge of the signal efficiency;
for the $\Vub$ values, it comes from the the theoretical uncertainty on $\Delta \zeta$.
For Ref.~\cite{blnp} we use the exponential parametrization
of the shape function.}
\vspace{0.1in}
\begin{tabular*}{\textwidth}{c@{\extracolsep{\fill}}ccccccc} 
\hline\hline\vspace{1mm}
  {\em Method} & $N_u$ & $N_u^{\rm{out}}$ & $BG_u$  & $\epsilon_\mathrm{sel}^u \epsilon_\mathrm{kin}^u$ &$\frac{\epsilon_\ell^\mathrm{sl} \epsilon_t^\mathrm{sl} } {\epsilon_\ell^u \epsilon_t^u }$  & $\Delta \BR(\Bxulnu)\ (10^{-3})$ & $\Vub\ \times(10^{-3})$       \\\hline\vspace{0.04in}
           &   &   &  &              &                 & & $4.27 \pm 0.16 \pm 0.13 \pm 0.30$~\cite{blnp} \\\vspace{0.04in}
 \rb{$\mX$}  & \rb{$803 \pm 60$}  & \rb{$27 \pm 2$} & \rb{$923\pm21$}& \rb{$0.331\pm0.003$} & \rb{$0.76\pm0.02$} &  \rb{$1.18 \pm 0.09 \pm 0.07 \pm 0.01$} &$4.56 \pm 0.17 \pm 0.14 \pm 0.32$~\cite{dge} \\\hline\vspace{0.04in}     
           &  & & &      &                             & & $3.88 \pm 0.19 \pm 0.16 \pm 0.28$~\cite{blnp} \\\vspace{0.04in}
\rb{$\Pplus$}   & \rb{$633 \pm 63$} & \rb{$48 \pm5$} & \rb{$1183\pm 27$} & \rb{$0.344\pm0.003$} & \rb{$0.81\pm0.02$} &  \rb{$0.95 \pm 0.10 \pm 0.08 \pm 0.01$} & $3.99 \pm 0.20 \pm 0.16 \pm 0.24$~\cite{dge} \\\hline\vspace{0.04in}
           &  &   &   &  &                      &   & $4.57 \pm 0.22 \pm 0.19 \pm 0.30$~\cite{blnp} \\\vspace{0.04in}
 $\mx, \Q$  & $ 562 \pm55$& $32\pm2$ & $789\pm9$ &$0.353\pm0.005$& $0.79\pm0.03$ &  $0.81 \pm 0.08 \pm 0.07 \pm 0.02$  & $4.64 \pm 0.23 \pm 0.19 \pm 0.25$~\cite{dge} \\\vspace{0.04in}
           & &     &  &  &                                &  & $4.93 \pm 0.24 \pm 0.20 \pm 0.36$~\cite{bll} \\
\hline\hline
\end{tabular*}
\label{tab:inputdeltaB}
\end{center}
\end{table*}

\begin{table*}[htbp]
\begin{center}
\caption{Contributions to the systematic uncertainty on the measured $\Delta \BR(\Bxulnu)$, shown in percent ($\%$)  for the three
kinematic cuts, from:
detector, shape function (input parameters and functional form), exclusive $\BR(\Bxulnu)$, gluon splitting,
exclusive $\BR(\Bxclnu)$, $\B\to\Dstar\ell^-\overline{\nu}$ form factors, $\BR(D)$, \mes\ fit, MC
statistics. The last column gives the total systematic uncertainty.
}
\vspace{0.1in}
\begin{tabular}{ccccccccccc}
\hline\hline
        &          &  \rbd{Shape}    &  \rbd{$\BR(\Bxulnu)$}  & \rbd{Gluon}     &               & \rbd{$\B\to\Dstar\ell^-\overline{\nu}$} &                       &        & \rbd{Monte Carlo}    &  \\ 
 \rb{{\em Method}} & \rb{Detector}  & \rbu{function} & \rbu{$X_u = \pi, \rho, \ldots$} & \rbu{splitting} & \rb{$\BR(\Bxclnu)$}  & \rbu{form factors}                      & \rb{$\BR(D)$} & \rb{\mes\ fit}  & \rbu{statistics}     &  \rb{Total}  \\\hline
\mx      &  1.92    & 0.90 &    2.08       &  1.62     &   0.87        &   0.21                  &   0.44   &  3.71   & 3.22  & 6.07  \\
\Pplus   &  3.88    & 1.31 &    2.22       &  1.47     &   2.80        &   0.39                  &   0.73   &  3.98   & 4.62  & 8.38  \\
\mX, \Q   &  3.83    & 2.43 &    2.71       &  1.02     &   1.17        &   0.55                  &   0.79   &  5.17   & 4.29  & 8.81   \\
\hline\hline
\end{tabular}
\label{tab:syst}
\end{center}
\end{table*}

We consider several sources of systematic uncertainties. Detector-related uncertainties 
take into account particle ($e$, $\mu$, $K$) identification (efficiency, mis-identification),
charged particle tracking efficiency, photon reconstruction efficiency and
\KL\ interactions.
We estimate the uncertainty due to signal and background modeling.
The uncertainty on the signal modeling are due
to the modeling of exclusive charmless semileptonic decays and gluon splitting into $s{\bar{s}}$-quark pairs.
We also calculate the uncertainties due to the non-perturbative parameters
and the functional form of the shape function.
The background simulation depends on the \B\ and $D$ branching fractions and 
$\Btodstlnu$ form factors; the corresponding systematic uncertainties are
calculated by varying all these quantities within their experimental errors.
We estimate the error due to \mes\ fits, coming from the uncertainty in the parameterization ansatz.
Finally, we estimate the error due to MC statistics.
The fractional contribution of each uncertainty is 
shown in Table~\ref{tab:syst} together with the total error.

The results of the partial branching fractions are translated into $\Vub$
in the context of recent QCD calculations~\cite{blnp,dge,bll}, including estimates
of theoretical uncertainties (see Table~\ref{tab:inputdeltaB}).
The hadronic input parameters, the $b$-quark mass $m_b$,  and the kinetic energy
expectation value $\mu_\pi^2$, are extracted from moment measurements in 
\Bsg and \Bxclnu. Their values in the kinetic scheme~\cite{kinetic} are
$m_b = (4.59 \pm 0.04)~\gevcc$ and $\mu_\pi^2 = (0.40 \pm 0.04)~{\rm GeV^2/c^2}$~\cite{buchmuellerflaecher}
and are translated into values in different schemes, as needed~\cite{blnp,dge,bll}.
The partial branching fraction $\Delta \BR(\Bxulnu)$ is related directly to 
$\Vub$ by the relation $\Vub = [\Delta \BR(\Bxulnu)/\tau_b\Delta \zeta ]^{1/2}$,
where $\tau_b$ is the average \B\ lifetime~\cite{PDG2006}, and $\Delta \zeta$ is the 
prediction for the partial rate for $\Bxulnu$ in the given phase-space region ~\cite{blnp,dge,bll}. 

In summary, we have measured the branching fractions for inclusive charmless semileptonic \B\
decays $\Bxulnu$ in three overlapping regions of phase space.
Relying on theoretical predictions, we extract values
for the CKM matrix element \Vub\ from our measured $\Delta\BR$.
 
We find that the determinations of $|V_{ub}|$ agree at 1~$\sigma$ level in the BNLP framework for
the \mx and combined (\mx,\Q) analyses. The analysis based on \Pplus differs from the two
others at a 2.5~$\sigma$ level, as indicated also by other experiments~\cite{bellevub2005}.
The \mx\ analysis captures the largest portion of phase space and gives 
the most precise determination of $|V_{ub}|$. Within their stated theoretical uncertainties, 
the results based on BLNP and DGE give consistent results.
The result, based on the hadronic mass spectrum, supersedes our previously published
measurement~\cite{vub2004}, reducing the relative uncertainty by 40$\%$.
These values are in good agreement with other inclusive $|V_{ub}|$ determinations
and they are somewhat higher, though compatible, than the results based on exclusive charmless
semileptonic decays~\cite{PDG2006}.

\label{sec:Acknowledgments}

We would like to thank the many theorists with 
whom we have had valuable
discussions, in particular J.~R.~Andersen, E.~Gardi,
B.~Lange, Z.~Ligeti, M.~Neubert and G.~Paz.
We are grateful for the excellent luminosity and machine conditions
provided by our \pep2\ colleagues, 
and for the substantial dedicated effort from
the computing organizations that support \babar.
The collaborating institutions wish to thank 
SLAC for its support and kind hospitality. 
This work is supported by
DOE
and NSF (USA),
NSERC (Canada),
CEA and
CNRS-IN2P3
(France),
BMBF and DFG
(Germany),
INFN (Italy),
FOM (The Netherlands),
NFR (Norway),
MIST (Russia),
MEC (Spain), and
STFC (United Kingdom). 
Individuals have received support from the
Marie Curie EIF (European Union) and
the A.~P.~Sloan Foundation.


\begin{thebibliography}{99}
\bibitem{hadronicsystem}
We indicate with $X$ the hadronic system in semileptonic \B\ decays. We use 
the notation $X_u$ and $X_c$ when referring, respectively, to  a charmless
and charmed hadronic system.
\bibitem{chconj}
 Charge-conjugate modes  are  implied throughout this letter, unless explicitly stated.
\end{thebibliography}
\end{document}